# The interaction between air ions and aerosol particles in the atmosphere

**K.L. Aplin and R.G. Harrison**

Department of Meteorology, University of Reading, PO Box 243, Earley Gate, Reading, Berkshire, RG6 6BB, U.K.

**Abstract**

Charged particles are continually generated in atmospheric air, and the interaction between natural ionisation and atmospheric particles is complicated. It is of some climatic importance to establish if ions are implicated in particle formation. Atmospheric ion concentrations have been investigated here at high temporal resolution, using Gerdien ion analysers at a site where synchronous meteorological measurements were also made. The background ionisation rate was also monitored with a Geiger counter, enabling ion production from natural radioactivity to be distinguished from other effects. Measurements at 1Hz offer some promise in establishing the atmospheric electrical influences in ionic nucleation bursts, although combinations of other meteorological factors are also known to be significant. High time resolution meteorological and ion measurements are therefore clearly necessary in advancing basic understanding in the behaviour of atmospheric aerosol.

**1. Introduction**

Natural radioactivity and cosmic rays are a constant source of molecular ions in atmospheric air. Establishing the significance of these long-lived atmospheric ions is intimately linked with the importance of the atmospheric electrical system in atmospheric processes (Harrison, 1997), and the conventional view that the electrical properties of the atmosphere are completely insignificant is increasingly untenable. The ions produced are rarely single species, but form clusters of water molecules around a central ion. Atmospheric electrical fields lead to transport of small ions by vertical conduction processes. Typical atmospheric ion concentrations in unpolluted air and fair weather conditions are about 400-500 ions cm$^{-3}$.

It is of great interest to understand how ions could be implicated in particle production, and significant ion-induced nucleation of particles could be a relevant mechanism. Hõrrak *et al.* (1998) reported the spontaneous formation of intermediate size ions in atmospheric air, from a change in the ion mobility spectrum of urban air. Observations of direct aerosol formation in unpolluted marine air (O'Dowd *et al.*, 1996) and theoretical work on clustering reactions of ions (Castleman, 1982) suggest ions can be critical in gas-to-particle conversion processes.

If the volumetric ion production rate is $q$, the positive ion number concentration $n_+$, the negative ion concentration $n_-$, and if it is assumed that $n_+ = n_- = n$, then the variation of ion concentration with time is given by the equation

$$\frac{dn}{dt} = q - \alpha n^2 - \beta nZ - \gamma n \qquad (1)$$

where $\alpha$ is the ionic self-recombination coefficient, $\beta$ the ion-aerosol attachment coefficient (which varies with aerosol size and charge), $Z$ the aerosol number concentration and $\gamma$ the ion-induced nucleation coefficient. It has been usual to regard the aerosol and ions as distinct systems of particles, which interact purely by collisions. However the observations of Hõrrak *et al.* (1998) now suggest that the $\gamma$-term could be significant in new particle events, and this would couple the time dependence in the ion concentration equations to time dependence in aerosol concentration. The simultaneous measurement of ion and aerosol concentration is therefore of importance.

## 2. The Gerdien Condenser

Gerdien developed a method of measuring air conductivity from a cylindrical condenser in 1905. A voltage is applied between two cylindrical electrodes and the inner one is earthed via a sensitive ammeter (Harrison, 1997). In the tube, ions of the same sign as the polarising voltage are repelled by the outer electrode, and move in the electric field to meet the inner electrode where they cause a small current. In the Gerdien used for this work (approximately 0.21 by 0.04m) the currents are of order $10^{-15}$ A, and the conductivity is given by

$$\sigma = \frac{i\varepsilon_0}{CV} \qquad (2)$$

where $C$ is the capacitance of the Gerdien, and $V$ is the polarising voltage across the electrodes. The Gerdien tube can be used to measure conductivity as long as the output current is proportional to the polarising voltage, indicating that a fixed fraction of the ions in the air are collected by the central electrode.

There exists a critical mobility $\mu_c$ such that only ions with $\mu > \mu_c$ contribute to the conductivity measurement. This is given by Wilkinson (1997)

$$\mu_c = \frac{(a^2 - b^2)\ln\left(\frac{a}{b}\right)u}{2VL} \qquad (3)$$

where $a$ is the radius of the tube, $b$ is the radius of the central electrode, $u$ is the flow speed through the tube and L is the length of the tube. Operating the Gerdien at different critical mobilities is easily executed by varying the voltage or flow rate, and enables it to act as a simple and effective mobility spectrometer.

## 3. Gerdien Control System

A Microcontroller-based system for remotely controlling and logging the Gerdien data has been developed. This is clearly necessary for measurement of mobility spectra, as the voltage across the Gerdien has to switch to different values. By this method, one Gerdien tube can also be used to measure quasi-synchronous positive and negative air conductivity by cycling between positive and negative polarising voltages.

The microcontroller used in this application[1] is programmable in BASIC and has eight I/O pins, which can be wired up for communication with other computers, or for digital inputs to external circuits. The microcontroller is cheap and compact, and can simultaneously control the Gerdien and send the data to the serial interface of another computer. Once the program is downloaded to the microcontroller it runs for as long it is powered, and the program resumes its operation if the power is cut and then restored.

---

[1] BASIC Stamp 1, Parallax Instruments Inc. 3805 Atherton Road, Suite 102, Rocklin, California 95765, USA

The electronic control circuit employed essentially comprises two relays, driven by VN10KM MOSFET transistors controlled by digital outputs from the microcontroller, and a LTC1298 12-bit analogue to digital converter. One relay (a changeover device) was connected to a positive and negative bias voltage. The other relay was a high input impedance reed relay, connected across the feedback resistor of the picoammeter. When this relay is on it shorts the picoammeter feedback resistor and provides a measurement of the picoammeter input offset voltage, overriding the state of the other relay so that no current measurement is made. The analogue-to-digital conversion is carried out by a subroutine in the microcontroller program (Parallax, 1998) and enables the voltage output from the picoammeter to be logged.

| Name | Actions |
|------|---------|
| Zero | Turns reed relay on to zero current amplifier |
|      | Sets other relay to "0" to reduce power consumption |
|      | Waits for 6s |
|      | Takes 5 readings of zeroed state of current amplifier at 1Hz |
|      | Sends readings to PC via serial protocol |
| +    | Turns reed relay off to allow ion current measurements |
|      | Sets other relay to positive bias voltage |
|      | Waits 3 minutes |
|      | Takes 10 readings of ion current at 1Hz |
|      | Sends readings to PC via serial protocol |
| -    | Sets relay to negative bias voltage |
|      | As + above and repeats forever |

Table 1: Outlines the actions performed by the program running on the microcontroller. The delay of three minutes after the change of bias voltage is required to allow complete recovery from the transient that occurs on switching.

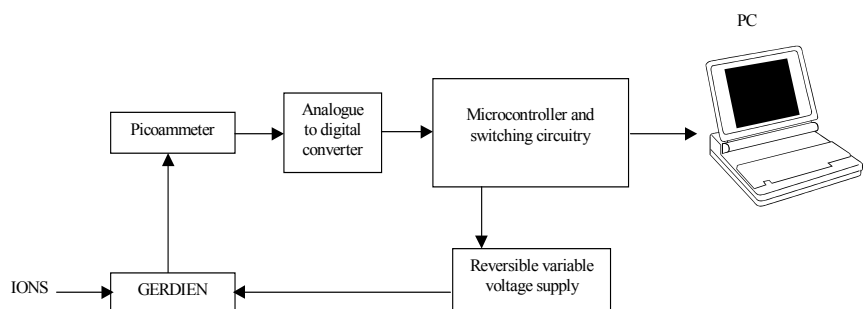

Figure 1: Schematic showing the integrated air ion measurement system

### 4. Results

Measurements of ion concentrations at different mobilities have been made in Reading by using the Microcontroller system to switch the Gerdien between two voltages of the same sign, –6V and –9V. Knowing the wind component into the Gerdien and the tube ventilation rate allows calculation of $\mu_c$ using (3). With only a fan controlling the flow into the tube ($u$=2.14ms$^{-1}$) these voltages correspond to critical mobilities of 3.1 and 2.1 x 10$^{-4}$ Vm$^{-2}$s$^{-1}$ respectively. In practice, the external wind component modulates the critical mobility and the mean value is reported below.

414

The ionic number density is approximated by

$$n = \frac{\sigma}{e\mu} \qquad (4)$$

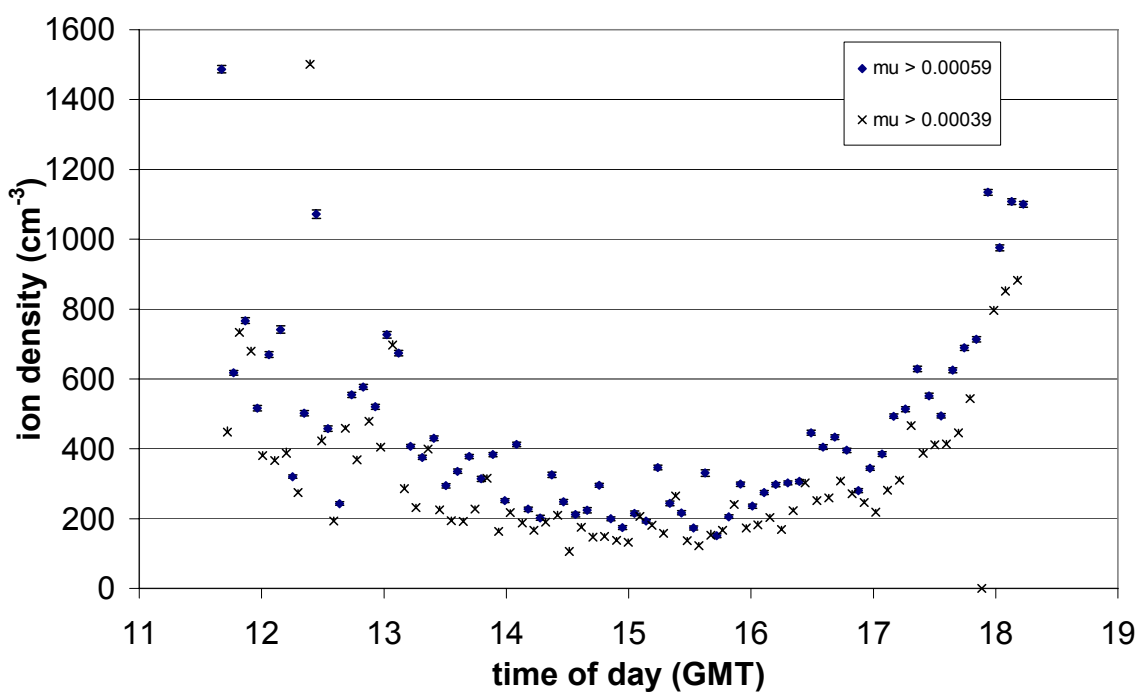

Figure 2: The negative ion number concentration at mean critical mobilities of 5.9 and 3.9 x $10^{-4}$ $Vm^{-2}s^{-1}$ for a typical day (23/2/99).

where $e$ is the electronic charge, $\mu$ the critical mobility and $n$ the concentration of negative ions with a mobility exceeding the critical mobility. Initial results are promising.

**5. Conclusions**

These results indicate that the microcontroller-based ion counter system is ideal for measuring high time and mobility resolution ion spectra, and the microcontroller system can easily be expanded, such as for controlling the fan. Synchronous measurement of mobility spectra and meteorological parameters will assist in the understanding of ion-aerosol interactions.

**References**

Castleman A.G. (1982), In: Schryer D.R., *Heterogeneous atmospheric chemistry*, AGU, Washington

Harrison R.G. (1997), Climate change and the global atmospheric electrical system, *Atmospheric Environment*, **31**, 20, 3483-3484

Hõrrak U. *et al.* (1998), Bursts of intermediate ions in atmospheric air, *J. Geophys. Res*, **103**, 13909-13915

O'Dowd C.D. *et al.* (1996), New particle formation in the environment, *Proc. 14th Int. Conf. on Nucleation and Atmospheric Aerosols,* Helsinki, Eds: Kulmala M. and Wagner P.E., Pergamon Press, 925-929

Parallax Inc. (1998), *BASIC Stamp ® Manual*, Version 1.9

Wilkinson S. (1997), *Determination of the characteristics of urban atmospheric aerosol,* MSc. Dissertation, Department of Meteorology, University of Reading